\documentstyle[aps,preprint,epsf]{revtex}
\tightenlines

\begin{document}
\title{Treating Some Solid State Problems with the Dirac Equation}
\author{R. Renan, M. H. Pacheco, C. A. S. Almeida}
\address{Universidade Federal do Cear\'{a}\\
{\it Physics Department}\\
{{\it C.P.} 6030, 60470-455}\\
Fortaleza-Ce, Brazil\thanks{%
Electronic address: renan@fisica.ufc.br, carlos@fisica.ufc.br}}
\maketitle

\begin{abstract}
The ambiguity involved in the definition of effective-mass
Hamiltonians for nonrelativistic models is resolved using the
Dirac equation. The multistep approximation is extended for
relativistic cases allowing the treatment of arbitrary potential
and effective-mass profiles without ordering problems. On the
other hand, if the Schr\"{o}dinger equation is supposed to be
used, our relativistic approach demonstrate that both results are
coincidents if the BenDaniel and Duke prescription for the
kinetic-energy operator is implemented. Applications for
semiconductor heterostructures are discussed.
\end{abstract}

\vspace{0.5cm}


\vspace{1.0cm}

The effective-mass theory has been successfully used in semiconductor
heterostructures \cite{vinter}. An interesting aspect arises when we treat
materials whose properties change from region to region. In particular, when
the effective mass depends on position, the Schr\"{o}dinger equation for an
arbitrary potential profile is usually solved numerically by different
methods. However, one of the problems of the effective-mass theory for
semiconductors heterostructures, is to decide how to write out the
Hamiltonian operator. This problem arise from canonical quantization of the
classical Hamiltonian. For position-dependent carrier effective mass, we
have an ordering problem with the kinetic energy operator (KEO). Some
authors proposed different forms for the kinetic energy operator, all having
the generic form proposed by von Roos~\cite{roos}
\begin{equation}
\hat{T}=\frac 14\left( m^\alpha (x)\hat{p}m^\beta \hat{p}m^\gamma
(x)+m^\gamma (x)\hat{p}m^\beta \hat{p}m^\alpha (x)\right) ,  \label{eqroos}
\end{equation}
where $\alpha +\beta +\gamma =-1$, but the problem was not resolved because
there is not a first principle to fix only one operator.

This ambiguity indicates that the Schr\"{o}dinger equation is not rigorously
suitable in the effective-mass aproximation with position-dependent
effective mass. It is reasonable to try other equation that represents the
same physics of the Schr\"{o}dinger equation in the low energy limit.

The object of this Letter is to demonstrate that, in fact, the Dirac
equation (at adequate limits) can successfully be used to describe
quantum-mechanical systems where position-dependent effective mass is
present. We recall that, in Dirac equation, the kinetic energy operator and
the mass term appear separately, so there is no ordering problems in this
context.

Obviously the considerations of this work only concern the
mathematical issues related with the equations of motion. Indeed,
a physically sensible application of the Dirac equation to
semiconductor heterostructures would have to take in to account a
relativistic extension of the Wannier-Slater theorem
\cite{landsberg}.

In this work we use a numerical method (multistep potential approximation
\cite{yto}) which has been applied to solve Schr\"{o}dinger equation for an
arbitrary potential profile. Here we extend this algorithm to the
relativistic case, in such a way that the ambiguity problem is overcome. In
particular, we consider the Dirac equation with a one dimensional arbitrary
potential well and find the energy levels for a particle. Also we apply the
method for a particular type of heterostructure and compare the results to
those obtained in the context of the Schr\"{o}dinger equation with the form (%
\ref{eqroos}) for the KEO and several choices for the parameters $\alpha
,\beta $ and $\gamma .$ For a KEO in the Schr\"{o}dinger equation of the
form suggested by BenDaniel and Duke ($\beta =-1,\gamma =0$) \cite{ben}, we
conclude that both equations lead to the same solutions in the energy range
concerned.

Let us now introduce the numerical method that allows us to obtain the
energy levels for a Dirac equation with space-dependent effective mass.

We will consider a particle with mass $m(z)$ that is submitted to an
arbitrary one dimensional potential well $V(z)$. The time-independent Dirac
equation is written as (in units with $\hbar =c=1$) \cite{ryder}
\begin{equation}
\left( \alpha \hat{p}+\beta m(z)\right) \Psi =\left( E-V(z)\right) \Psi ,
\label{dirac}
\end{equation}
where, $\hat{p}=-i\displaystyle{\frac d{dz}}$ is the momentum operator, $E$
is the electron energy, $\alpha $ and $\beta $ are $4\times 4$ matrices
given by
\begin{equation}
\alpha =\left(
\begin{array}{cc}
0 & \sigma ^3 \\
\sigma ^3 & ~0
\end{array}
\right) ,\quad \quad \quad \beta =\left(
\begin{array}{lr}
I & ~0 \\
0 & -I
\end{array}
\right) ,  \label{dmatrix}
\end{equation}
$I$ is the $2\times 2$ identity matrix and $\sigma ^3$ is a $2\times 2$
Pauli matrix defined as
\begin{equation}
\sigma ^3=\left(
\begin{array}{lr}
1 & ~0 \\
0 & -1
\end{array}
\right)  \label{pauli}
\end{equation}

Consider now an arbitrary well as sketched in fig.1. We split up the
interval $[a,b]$ into $N$ infinitesimal intervals of length $\Delta
z=(b-a)/N $. For the $i$-th interval, we approximate the potential and the
mass by
\begin{equation}
V(z)=V(z_i)=V_i\quad and\quad m(z)=m(z_i)=m_i,\quad \quad for\quad z_i\le
z<z_{i+1}.  \label{masspot}
\end{equation}
The wave function of the electron in Dirac's equation with no spin
flip in the $i-$th interval is
\begin{equation}
\psi _i\left( z\right) =A_ie^{ip_iz}\left( \!\!
\begin{array}{c}
1 \\
0 \\
\displaystyle{\frac{p_i}{E-v_i+m_i}} \\
0
\end{array}
\!\!\right) +B_ie^{-ip_iz}\left( \!\!
\begin{array}{c}
1 \\
0 \\
\displaystyle{\frac{-p_i}{E-v_i+m_i}} \\
0
\end{array}
\!\!\right) ,  \label{diracsol}
\end{equation}
where $p_i=\sqrt{\left( E-V_i\right) ^2-m_i^2}$. The bound state
conditions are given by $|E-V_0|<m_0$ and $|E-V_{N+1}|<m_{N+1}$.
By imposing the continuity of the wave function at each $z=z_i$,
we have a matrix $M(E)$ which relates the coefficients in the
region where $z<a$ with the region where $z>b$.
\begin{equation}
\left(
\begin{array}{c}
A_{N+1} \\
B_{N+1}
\end{array}
\right) =M(E)\left(
\begin{array}{c}
A_0 \\
B_0
\end{array}
\right)  \label{matrix}
\end{equation}
The finiteness of the wave function requires that
\begin{equation}
M(E)_{22}=0  \label{wfinite}
\end{equation}

So, the solution of ~(\ref{wfinite}) give us the energy levels.

Note that this numerical method is specially convenient for treating wells
and barriers with arbitrary profiles and it is nothing else than an
extension of the transfer-matrix method for relativistic theories. To the
best of our knowledge, this is the first numerical analysis for evaluating
energy levels in an relativistic equation with mass position-dependent and
arbitrary potential. However, our main point here is to demonstrate that the
Dirac equation can be used to obtain unambiguously results in situations
where the Schr\"{o}dinger equation depends on ordering problems.

As an illustration, we applied the method described above for an electron in
a one-dimensional GaAs/Al$_{0.3}$Ga$_{0.7}$As heterostructure. For the sake
of comparison with previous results we take a square well, as sketched in
fig. 2. The electron effective mass is $0.67m_0$ and $0.86m_0$ for GaAs and
Al$_{0.3}$Ga$_{0.7}$As respectively~( $m_0$ is the free electron mass)(Fig.
2). In the Schr\"{o}dinger equation we use the von Ross operator~\cite{roos}
considering several values of the $\alpha $ parameter. Note that for abrupt
heterojunctions only Hamiltonians with $\alpha =\gamma $ are viable, due to
continuity conditions across the heterojunction \cite{morrow}. As we can see
in fig. 3 there is an extraordinary coincidence between the results from the
Dirac and from the Schr\"{o}dinger equations for $\alpha =0$. As a matter of
fact, this is expected since the maximum value of the energy involved is
3eV. Further, this result strongly supports the prescription of BenDaniel
and Duke \cite{ben} for the Schr\"{o}dinger context.

As a second illustration consider the conduction-band structure of a GaAs/Al$%
_{0.3}$Ga$_{0.7}$As system with nonabrupt interface and assume that the
effective mass changes linearly at the transition regions while the
potential well varies quadratically in that regions (denoted by $a$) as it
is shown in fig. 4. There the potential is given by

\begin{equation}
V(z)=C\left[ \epsilon _1\chi (z)+\epsilon _2\chi (z)^2\right] \text{ ,}
\label{potential}
\end{equation}
where $C=0.6$ is the conduction band offset and $\epsilon _1=0.3$, $\epsilon
_2=0.7$ are constants associated with the compositional dependence of the
energy-gap difference between GaAs and AlGaAs (experimental parameters and
details concerned can be seen in refs. \cite{renan,adachi}).

Again, we use our relativistic method and the Schr\"{o}dinger equation with
the BenDaniel and Duke prescription ($\alpha =\gamma =0$). Once more, as we
can see in fig. 5, a complete coincidence between the relativistic and
non-relativistic results is obtained.

In conclusion, we have shown that a relativistic method can be
successfully used to overcome the ordering problem of the kinetic
energy operator in non-relativistic models. Since the range of
energy involved is extremely low (comparing to electron rest
mass), the numerical results are perfectly coincident in both
cases.

It is worthwhile to mention that, notwithstanding the
Wannier-Slater theorem commented in the introduction, we claim
attention for the coincidence shown above. Therefore, we believe
that, after exhaustive and appropriate considerations about
effective-mass approximation, band structures and the periodic
potential, the relativistic approach constructed here it could be
used to calculate physical parameters in the theory of abrupt and
nonabrupt semiconductor heterostructures.

Moreover, the relativistic treatment developed here can be applied to all
physical systems described by a Sturm-Liouville eigenvalue equation (within
an appropriate range of energy), namely

\begin{equation}
-\frac d{dz}\left[ \frac 1{p(z)}\frac{df}{dz}\right] +q(z)f(z)=\lambda
w(z)f(z)\text{ }.  \label{sturm}
\end{equation}
Here $f$ is an eigenfunction, $\lambda $ is an eigenvalue and $p$,
$q$, and $w$ describe particular properties of the system. For
example, this equation can describe the motion of electrons or
phonons along the growth axis of a [100] zinc-blende
heterostructure \cite{foreman}. Thereby, the equation
(\ref{sturm}) can always be replaced by an correspondent Dirac
equation for avoid ordering problems.

{\bf Acknowledgments} 

M. H. Pacheco was supported in part by Conselho Nacional de
Desenvolvimento Cient\'{\i}fico e Tecnol\'{o}gico-CNPq.

\begin{center}
{\bf FIGURE CAPTIONS}
\end{center}

Figure 1: Generic profile of a one-dimensional quantum potential well. We
split up the interval $[a,b]$ into $N$ infinitesimal intervals of length $%
\Delta z=(b-a)/N$

Figure 2: Potential well and effective mass, as a function of the position,
for an abrupt heterostructure.

Figure 3: Eigenenergies in the conduction band of the GaAs/Al$_{0.3}$Ga$%
_{0.7}$As, with a conduction band-offset of $0.6$ vs the well width (abrupt
heterojunction). The chosen value for $a$ is $20\%$ of the well width. Solid
line shows the calculations performed using the Dirac equation. Dotted line
shows the calculations for Schr\"{o}dinger equation using the BenDaniel-Duke
prescription ($\alpha =0$). The + + curve denotes calculations using the Zhu
and Kroemer prescription ($\alpha =-0.5$).

Figure 4: Potential well and effective mass, as a function of the position,
for a nonabrupt heterostructure ( $a$ denotes the transition region).

Figure 5: Eigenenergies in the conduction band of the GaAs/Al$_{0.3}$Ga$%
_{0.7}$As, with a conduction band-offset of $0.6$ vs the well width
(nonabrupt heterojunction). The chosen value for $a$ is $20\%$ of the well
width. Solid line shows the calculations performed using the Dirac equation
and dotted line shows the calculations for Schr\"{o}dinger equation using
the BenDaniel-Duke prescription ($\alpha =\gamma =0$). The + + curve denotes
calculations using the Zhu and Kroemer prescription ($\alpha =\gamma =-0.5$).


\begin{references}
\bibitem{vinter}  C. Weisbuch and B. Vinter, {\it Quantum Semiconductor
Heterostructures }(Academic Press, New York, 1993)

\bibitem{roos}  O. von Roos, Phys. Rev. {\bf B 27}, 7547 (1983)

\bibitem{landsberg}  P. T. Landsberg, {\it Solid State Theory: Methods and
Applications }(Wiley-Interscience, London, 1969)

\bibitem{yto}  Y. Ando and T. Itoh, J. Appl. Phys. {\bf 61}, 1497 (1987)

\bibitem{ben}  D. J. BenDaniel and C. B. Duke, Phys. Rev.{\bf \ 152}, 683
(1966)

\bibitem{renan}  R. Renan, V. N. Freire, M.M. Auto and G. A. Farias, Phys.
Rev. {\bf B 48}, 8446 (1993)

\bibitem{adachi}  S. Adachi, J. Appl. Phys. {\bf 58}, R1 (1985)

\bibitem{ryder}  L. H. Ryder, {\it Quantum Field Theory }( Cambridge
University Press, Cambridge, England, 1995)

\bibitem{morrow}  R. Morrow and K. Brownstein, Phys. Rev.{\bf \ B30}, 678
(1984)

\bibitem{foreman}  B. A. Foreman, Phys. Rev. Lett.{\bf \ 80}, 3823 (1998)

\newpage%
{}
\end{references}
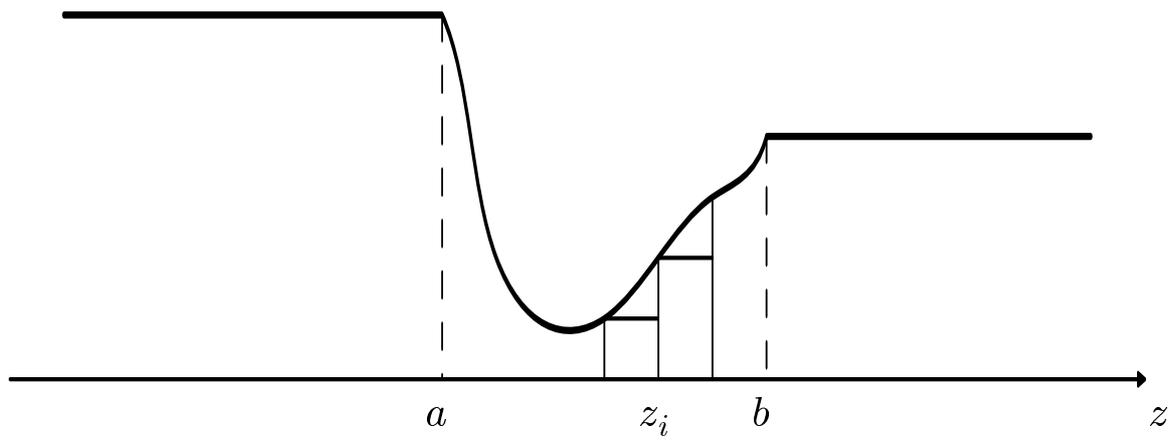
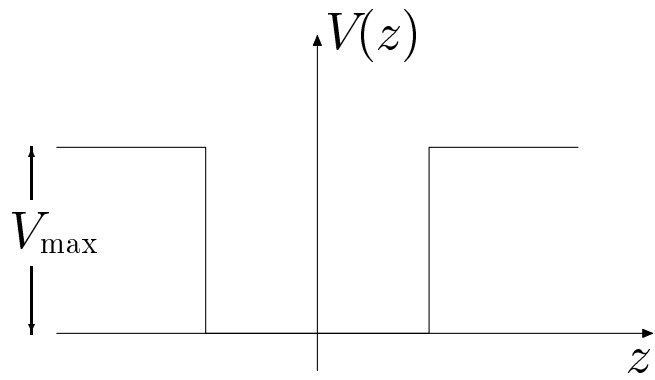
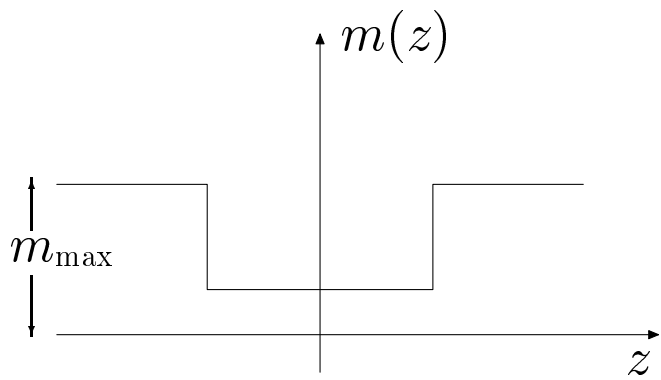
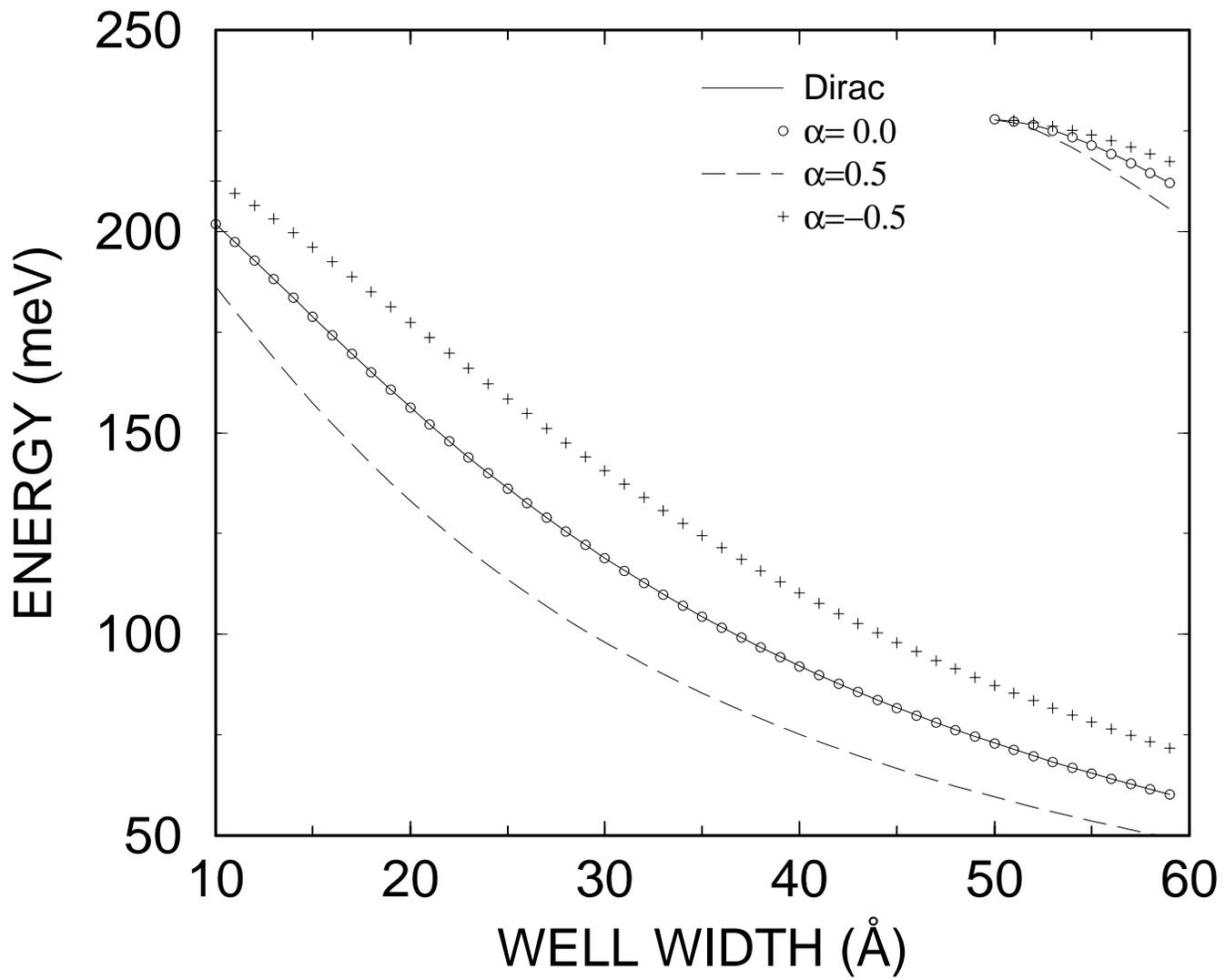
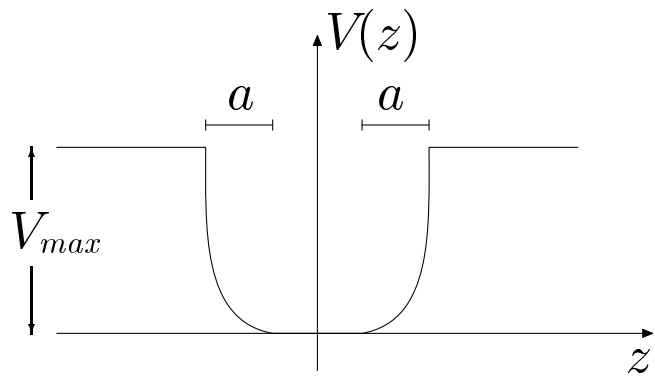
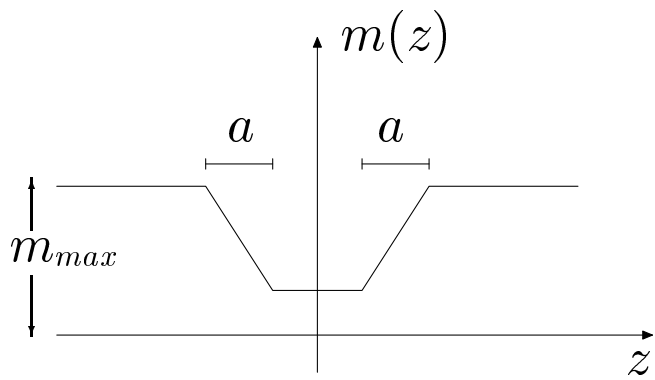
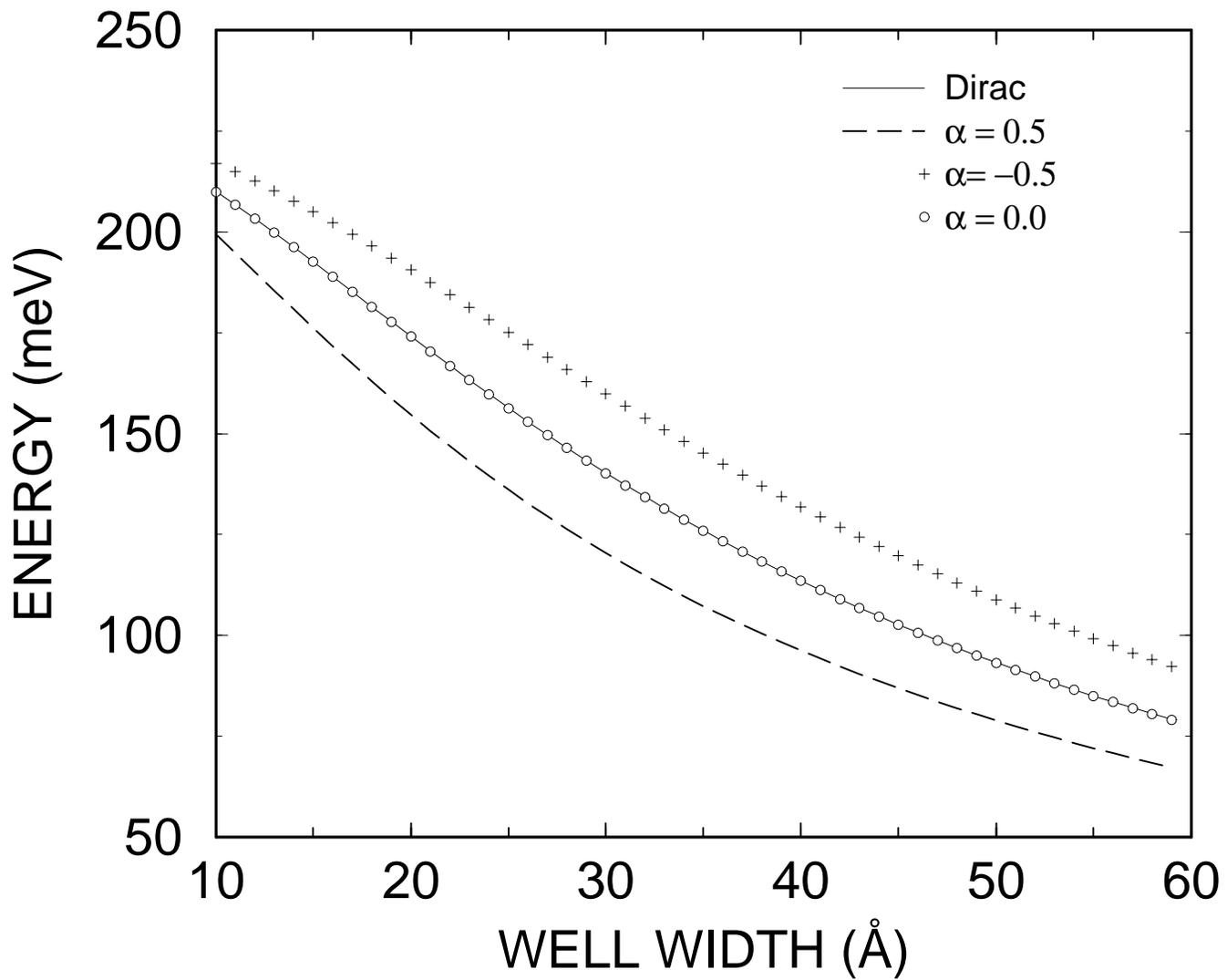
\end{document}